\documentstyle[12pt,aasms4]{article}
\def\lapprox{\hbox{\lower .8ex\hbox{$\,\buildrel < \over\sim\,$}}}
\def\gapprox{\hbox{\lower .8ex\hbox{$\,\buildrel > \over\sim\,$}}}

\begin{document}

\title 
{Tycho Brahe's supernova: light from centuries past}

\bigskip

\author
{Pilar Ruiz--Lapuente \altaffilmark{1,2}}

\altaffiltext
{1}{ Department of Astronomy, University of Barcelona, Mart\'\i\ i Franqu\'es
1, E--08028 Barcelona, Spain. E--mail: pilar@am.ub.es}

\altaffiltext
{2}{Max--Planck--Institut f\"ur Astrophysik, Karl--Schwarzschild--Strasse 1,
D--85740 Garching, Federal Republic of Germany. E--mail:
pilar@MPA--Garching.MPG.DE}

\slugcomment{{\it Running title:} Tycho Brahe supernova}

\begin{abstract}

The light curve of SN 1572 is described in the terms  
used nowadays to characterize SNeIa.
By assembling the records of the observations done in 1572--74
and evaluating their uncertainties, it is possible to
recover the light curve and the color evolution
of this supernova. It is found that, within the SNe Ia family,  
the event should have been a SNIa with a normal
rate of decline, its stretch factor being {\it s} $\sim$ 0.9. 
Visual light curve near maximum,
late--time decline and the color evolution  sustain
this conclusion. After correcting for extinction, the luminosity 
of this supernova is found to be 
M$_{V}$ $=$ --19.58 --5 log (D/3.5 kpc) $\pm$ 0.42.

\end{abstract}


\bigskip

\section{Introduction}

SN 1572 was well observed in Europe (as well as in the Far--East) 
for almost two years.   
It added a new aspect to the debate at the time over 
the Aristotelian cosmological views, as it forced to reconsider the
immutability of the heavens and the solid nature of the
celestial spheres: the ``star'' gained
brightness and lost it during a period of two years, but it showed  
no detectable parallax. 
According to prevalent views about the heavens,
mutability would only happen in the sublunar region. This was even the
place where comets were assumed to originate. The appearence of SN 1572
challenged the order of the celestial spheres. The observers
who followed it up\footnote{The observers who mostly contributed to measure
the position and luminosity,
 Tycho Brahe, Thomas Digges, Thaddeus Hagecius, 
Michael M\"astlin, Jer\'onimo Mu\~noz, 
Caspar Peucer \& Johannes Pr\"atorius held very different views on
the meaning of SN 1572.  
A comparison of their measurements and an account of
 their views is given elsewhere 

\noindent
(Ruiz--Lapuente 2003).}
took sides with respect to established Aristotelian
views. Today we can
still use their observations to see whether that supernova would be
of use for cosmology in a different way, as a distance indicator if
seen by observers billions of years away from us. It is indeed 
possible to have a clear idea of where SN 1572 stands among its class.

After several centuries of questioning the nature of SN 1572, 
the identification as a Type I supernova came through
the revision of the light curve done by Baade (1945) and based on  
data taken or quoted by Tycho Brahe (1603a). 
Before that time there were still speculations on whether it was a
variable star of some kind, a nova (Morgan 1945) 
and it was even still considered the suggestion of 
its cometary nature (Lynn 1883). The cometary idea expressed in 1573 by 
Jer\'onimo Mu\~noz was based on the fact that the event is 
aligned with the Milky Way, so 
its decay in luminosity could be explained if it was a
comet born among the stars that would first  
approach and then move away from us just along the line of sight. 
 Tycho's {\it stella nova}
 lies indeed only  49--98 pc above the Galactic plane.

In modern times, its comparison with other SNe allowed its classification
as a Type I supernova (Baade 1945). Later on, this class was shown to 
contain events of very different nature: those identified
with the explosion of white dwarfs (Type Ia) and those corresponding to the
collapse of massive stars whose envelope had lost its  
hydrogen content in the interaction with a binary companion (Type Ib or c).  
SN 1572 was of the Type Ia class as discussed by de Vaucouleurs
(1985) and by van den Bergh (1993). No further doubts that it is 
a Type Ia SN can now 
be held in view of the X--ray spectrum, which clearly differentiates
SNIa events from SNe coming from the collapse of massive stars 
(Hughes et al. 1995). A first comparison of its luminosity at maximum
with the bulk of SNeIa 
would have led to think that it was fainter than normal SNeIa. 
van den Bergh (1993) considers whether it could be a peculiar, 
subluminous SNIa, like SN 1991bg. However, SN 1572 was heavily obscured.
It was reddened by 
E(B--V)=0.6$\pm$ 0.04 as it corresponds to the reddening of the stars 
near its position (Ruiz--Lapuente et al. 2003a,b).
After taking into account 
the extinction undergone across the Milky Way, as well as the
measurement of its decline rate, it is concluded that SN1572 was not a 
91bg--like event. Neither was it 
an overluminous SNIa like SN 1991T or similars, but
rather an event in the middle of the SNIa class.

\section{New  evidence on visual, color evolution and late decline} 

\subsection{ Visual light curve till 60 days} 

The date of maximum of SN 1572 is, from the available accounts, 
the most uncertain record. The first observation was done on Nov
6 though there is no record of the magnitude at that date
(see for a full discussion 
Stephenson \& Green 2002). On Nov 2 it was not yet noticeable 
as reported by Mu\~noz
(1573). In Nov 11, date of the first observation done by 
Tycho Brahe (1603a) and the first reported observation by Mu\~noz
(1573), 
it seems to have achieved  a magnitude between those of Jupiter and
Venus. The Nov 5 detection given in 
the report of the {\it Progymnasmata} 
by Tycho Brahe (1603b) is an erroneous date, as we have checked in
the original. 
On Nov 16 and 17, when Caspar Peucer and Johannes Pr\"atorius
saw it, it seems to have a magnitude close to Venus (Table 1 and notes). 
 Later on, in
Jan 7, it was already fainter than Jupiter (Mu\~noz 1573).  
The reconstruction of the observations suggests 
an apparent visual magnitude at maximum 
of --4.0$\pm$ 0.3 (Baade 1945; de Vaucouleurs 1985). 
The overall light curve fitting done in the present work by 
trying different decline rates (see below) indicates that the maximum 
was broad
and that it should have taken place around Nov 21.
On Nov 11, and even on
Nov 16, when Peucer and Pr\"atorius 
observed it, it would be still on the rise.   
This is consistent with the mean risetime to maximum 
of SNeIa of 17.8$^{+1.9}_{-1.0}$ days (Goldhaber et al. 2001).
While most authors have taken arbitrarily Nov 15 as the date
of visual maximum, it seems obvious that 
the maximum could not have happened before Nov 20.

The SN 1572 data are compared in this section 
to templates using the stretch factor {\it s}
as characterization of the rate of decline (Perlmutter et al. 1999; 
Goldhaber et al 2001; Nobili et al. 2003). 
The stretch factor {\it s} method is used by the Supernova Cosmology
Project to quantify the decline
rate of the supernova from data extending up to 60 days after maximum. Even 
in absence of a measurement of the brightness at maximum, the method
allows to produce a fine description of the supernova within the 
family of decline rates. 
The magnitudes and limits on magnitudes for SN 1572, when compared to
templates, lead us to conclude that this supernova was not fast.  
The best agreement found for SN 1572 corresponds to
  {\it s} $\sim$ 0.9. In this regard, 
SN 1572 is very similar to SN 1996X which had a {\it s} = 0.889. 
A comparison can be seen in Figure 1, where the data of SN 1572
are shown overplotted to a s = 0.9 SNIa template in V. We also
show for comparison the visual light curves
of SN 1996X (normal SNIa of {\it s} $=$ 0.889) and SN 1991bg
(a fast subluminous event of {\it s} $=$ 0.62). 
Therefore, these first 60 days show that there is no
similarity with the fast subluminous SN 1991bg.  
A broad light curve as the one of SN 1991T can also be excluded from
the constraints about its decline (see Figure 1).

\subsection{Late--time decline}

An additional proof
 that SN 1572 is a standard SNIa comes 
from its slow late decline between 100 and 450 days 
after maximum, similar to that of the bulk of SNeIa.
 SN 1572 had a late decline
of 1.4 mag in 100 days during that epoch,
 in concordance with normal SNeIa and
far from the fast decline of SN 1991bg (Filippenko et al. 1992; 
Leibundgut et al. 1993; Ruiz--Lapuente et al. 1993; 
Turatto et al. 1996).  SN 1990N and other normal SNeIa   
had decline rates in such period of
1.38--1.5 mag in 100 days.
 In consistency with  
the findings near maximum, late--time and overall 
similarity to the light curves of normal SNIa is found. 
In order to test the adequacy of the fit using the whole light curve,
 we construct new templates for a normal event of s = 0.9 
(well exemplified by SN 1996X), a
subluminous 91bg--type event and an overluminous
 91T--type event matching the available early and
late--time observations. Before 60 days those templates
are equal to the Hamuy et al. 1996b templates and the SCP templates for
the corresponding stretches of those SNeIa. At later times they follow
the available late--time photometry (Schmidt et 
al. 1993; Salvo et al. 2001) to their latest
available epoch.  
The fit of the SN 1572 data to the SNIa template corresponding to s =
 0.9 has a $\chi^{2}$ of 14.44 for 10 degrees
of freedom,  which is acceptable. In contrast, the fit to the template of a
fast declining, underluminous SNIa like SN 1991bg (s = 0.62) has an
exceedingly high $\chi^{2}$ of 53.55 for 4 degrees of freedom (the two
premaximum points as well as the last 4 points 
have been excluded because of the lack of data for
those subluminous SNeIa at such stage). In the other extreme, the fit to
the template of a slow-declining, overluminous SNIa like SN 1991T 
(s = 1.2) is almost as bad, with a $\chi^{2}$ of 82.99 for 10 degrees of
freedom.

In Figure 2, we show the overall 
visual light curve till almost 500 days compared with that of 
SN 1996X and the templates for SN 1991bg and SN 1991T. 
The last upper limit given by Tycho Brahe indicates 
that the supernova had not yet leveled off at 480 days past maximum
 due to a light echo produced in an intervening dust cloud. 
Not all the SNeIa in dusty regions have shown  light echoes
associated with the 
scattering of the supernova light in neighboring dust clouds, though. 
The production of light echos depends on the spatial distribution of
dust relative to both observer and supernova. In the case of SN 1986G, 
a very heavily reddened supernova, such light echo was not observed
(Schmidt et al. 1993). 
However, for SN 1572 one can not exclude a 
leveling off occuring later, at a fainter magnitude level.
SN 1991T and SN 1998bu  have indeed shown a slowing down 
of the V magnitude decline at around 400 days as result of the light
echo (Schmidt et al. 1993, Cappellaro  et al. 2001), and the leveling off
occurs at 500 days about 10 magnitudes below the maximum brigthness. 
 For SN 1572 an equivalent step in the level of brightness
would correspond to V$=$6, which is about 
the naked eye limit. 
SN 1998bu is a SNIa very
similar to SN 1996X and SN 1572 but with a 
reddening of E(B--V)$=$ 0.32 $\pm$ 0.04 
(Hernandez et al. 2000; Cappellaro et al. 2001). In fact, we can 
think of Tycho Brahe' s supernova as SN 1998bu  with twice the 
reddening.

Moreover, the late--time decline 
of SN 1572 informs us of
deposition of the energy
from the decay $^{56}$Co $\rightarrow$ $^{56}$Fe in the ejecta of
this supernova. Its slow decline would fit  
the declines predicted for SNeIa coming from the explosion 
of Chandrasekhar--mass WDs.
The late--time decline in V indicates a significant deposition of
the energy of the positrons which arise in the decay $^{56}$Co
$\rightarrow$ $^{56}$Fe. After 200 days since explosion, 
positrons become the dominant source of luminosity. 
Departures from full trapping of 
$^{56}$Co decay are seen in all SNeIa 
and they are expected. They arise from lack of confinement of positrons
to the ejecta of the supernova
 and/or incomplete thermalization of their energies.
Various effects can lead to the diversity in the late--time SN bolometric
light curves: nucleosynthesis, kinematic differences, ejected mass, 
degree of mixing of the ejecta, and configuration and intensity of the
magnetic field.  In particular, as suggested in
Ruiz--Lapuente \& Spruit (1998), those late--time declines can serve 
to investigate the magnetic field configuration of the SN and trace it
back to the original configuration of the WD right before explosion.
It is found that departures from the full
trapping of  $^{56}$Co decay of the order of 10--15$\%$ at about 400
days can be explained by the distribution of radioactive material.
Larger departures, of 30--40$\%$ or larger, have to be interpreted
in terms of lack of magnetic field confinement of positrons, or even
enhancement of the escape favored by a radially combed configuration of the
magnetic field. In the case of SN 1991bg the very fast drop of the late
light curve suggests not only a total ejected mass well below the 
Chandrasekhar mass, but an absence of a significant tangled magnetic
field. An exploration of the late--time decline
predicted for different explosion models 
compared with a sizeable sample of bolometric
light curves was done by Milne, The and Leising (1999). 
For SN 1572, the availability of data is limited to the visual band.
Though, analogies can be drawn with similar SNeIa for which the
bolometric data are available.

The late decline is, as seen above, a major proof which confirms
 that this historical supernova
is not part of the estimated 16$\%$ of intrinsically subluminous
SNeIa (Li et al. 2001), likely arising from 
a different class of explosion ejecting a smaller amount of mass
(Ruiz--Lapuente et al. 1993). If it were part
 of the class of peculiar
subluminous SNeIa, caution should be given when used for cosmology, because 
it is not clear whether those subluminous events are in the
linear relation brightness--rate of decline. 

\subsection{Reddening and color}

The most direct estimate of the reddening comes from our measurement 
of the reddening and extinction of the stars 
around the geometrical center of SN 1572 at the distance 
of the supernova (Ruiz--Lapuente et al. 2003a,b). The stars which are
compatible with a distance in the range of 
3$\pm$1 kpc (see de Vaucouleurs 1985
 for a review of the distances towards SN 1572) have reddenings which 
average to E(B--V)=0.6 $\pm$ 0.04. 
Those values were derived  
in the programme of search
of the companion star of Galactic SNeIa
(Ruiz--Lapuente et al. 2003a,b). Star candidates
within a radius compatible with the degree of precision on the
center and distance gained during the 431 years since explosion are
modeled to determine T$_{eff}$, log g, metallicity, distance and
E(B--V). The radial velocities are measured.
The individual reddening values 
found in that programme for the stars 
within 35$\%$ of the radius of the remnant
go from  E(B--V) = 0.50 to E(B--V)=0.8, the values increasing
with distance.

The above mean reddening to SN 1572 corresponds to an extinction 
$A_{V} =$ 1.86 $\pm$ 0.12 \ mag if we take $R_{V}=$ 3.1 (Sneden et
al. 1978; Riecke \& Lebofsky 1985). 
The Galactic extinction data based on 
COBE/DIRBE observations (Drimmel \& Spergel 2001)
 give a value of $A_{V} = 1.77 \ mag$ in that 
direction (the maximum Galactic extinction being $A_{V} = 1.90\ mag$
there). Thus the extinction measured at the distance of SN 1572 
is in agreement with the COBE results. 
Correcting the apparent brightness from the dimming by dust,
we find again that Tycho's SN was not subluminous but a ``normal'' SN Ia.

We can correct as well the color of the supernova from the records
of the epoch and derive the intrinsic color evolution. SNeIa 
are found to show a similar color evolution two months after
maximum, with a well--established law of low intrinsic dispersion valid
from 30 to 90 days, explored in Phillips et al. 1999 (Ph99) based on the work
of Lira (1995):

$$(B-V)_{0}=0.725 (\pm 0.05) -0.0118(t_{V}-60)  \eqno{(1)}$$    
 
\noindent
where $t_{V}$ is the time since visual maximum.

The color of SN 1572 two months after discovery 
was reported to be similar to Mars and
 Aldebaran (thus B--V in the range 1.36--1.54). Previous and
subsequent color estimates are also shown in Table 2. After 
correcting  the observed color by our measurement of 
reddening $E(B-V) =$ 0.6 $\pm$ 0.04, the intrinsic color at 55 $\pm$10
 days is 
$(B - V)_{0} =$ 0.76 $\pm$ 0.24, very much in agreement with the expected
$(B - V)_{0} =$ 0.78 $\pm$0.15 for that epoch. Such color fits very well
 in the  Ph99 law  
for the time evolution of the intrinsic color of the SNeIa. 
In agreement with that law, 
Nobili et al. (2002) have shown
that the color evolution of the bulk of SNeIa
has a low dispersion of 0.1 mag in the tail.
Before maximum the corrected color of SN 1572 would be
$(B - V)_{0} =$ 0.22 $\pm$ 0.29, thus 
consistent with the fact that normal SNeIa have $(B-V)_{0}$ $\sim$ 0. 
SN 1991bg and other subluminous SNeIa 
clearly deviate from the standard color evolution at early epochs,
being intrinsically red at maximum $(B-V)_{0}=0.6$ (Leibundgut et
al. 1993; Ph99). 
Moreover, the color evolution derived from the
B and V templates for a s $=$ 0.9 SNIa gives a good agreement with
the supernova. In Figure 3 we show the color light curve of SN 1572, corrected
by its E(B--V)$=$ 0.6, compared  
with that of ``normal'' SNeIa. Both the peculiar SN 1991bg
and the normal SN 1996X are also shown in their reddening--corrected
colors.

The data displayed in this figure come from  
different observers (see Table 2 and notes). Color estimates 
by eye are difficulted by 
the lack of an established system which encompases the whole range of
colors. The conversion of eye--estimated 
colors to our filter scale is addressed in Pskovskii (1977) and 
Schaefer (1996). To add to the inherent eye uncertainty in estimating
color, it has to be noted that  
the first comparisons done by the observers of the time 
using planets like Jupiter, Venus 
or Saturn as references, were difficulted by the fact that
the planets were often very far from the position
of SN 1572 (away in the sky
by 59$^{0}$).
De Vaucouleurs (1985) discussed the effect of
that on the early  visual estimates (later estimates since they are
done through comparison with particular stars are more reliable).

In the nebular phase, at 175 days according to Tycho Brahe's records 
(Table 2) the supernova went back to a white color, which is consistent, 
after correcting for extinction, with what is 
found in normal SNeIa as well. Uncertainties in the color estimates
have been assigned and  new records included.

\subsection{The luminosity of Tycho's supernova}

Given the apparent visual magnitude, the luminosity of Tycho Brahe's  
supernova would 
have corresponded to a visual absolute magnitude of 
 $M_{V}= -17.72 - 5\ log(d/3.5\ kpc) - A_{V}$. 
If corrected for $A_{V}$ =  1.86 $\pm$0.12, this means 
$M_{V}= -19.58 -5\ log (d/3.5\ kpc) \pm 0.42$. 

De Vaucouleurs (1985) gives 
as the most likely estimate to the distance to SN 1572
 3.2 $\pm$ 0.3 kpc. 
Taking 3.2 kpc for the distance,
 the luminosity would be --19.38 $\pm$ 0.42. That
compares well with the value $M_{V} = -19.12\pm0.26$, the mean magnitude 
from the Cal\'an/Tololo sample (Hamuy et al. 1996a).

To get a more refined distance to SN 1572, one would need to
follow a similar procedure as in Winkler, Gupta \& Knox
(2003) for SN 1006. These authors obtained a distance to SN 1006
of 2.18 $\pm$ 0.08 kpc  using the SNR expanding parallax method
(Kirshner, Winkler \& Chevalier 1987) with precise
measurements of the proper motion of the filaments in SN 1006
and narrower estimates of the shock velocity of the expanding
remnant. 

The proper motion of the filaments in SN 1572 was measured by 
Kamper \& van den Bergh (1978), and it is about 0.22 $\pm$ 0.02
 $^{''}$ yr$^{-1}$. There is, however, a larger uncertainty in the
estimate of the shock velocity, which is obtained through
 modeling of the broad H$\alpha$ emission 
of the material swept up by the SN.  The estimate is model
dependent and improves with a better knowledge of the composition
of the fully ionized plasma behind the shock. Kirshner, Winkler
\& Chevalier (1987) had suggested from the observations available
at that time a distance to SN 1006 between 1.4 and 2.1 kpc and 
a distance to SN 1572 of 2.0--2.8 kpc. The estimates of the shock velocity,
though, could soon be revised from the wealth of spectral observations
on SN 1572's filaments taken during the last years.

A rough
 estimate of the distance to Tycho's SN can be made by comparison of the
angular diameters of the two historical SNeIa.
 The angular diameter of G327.6+14.6 (the remnant of SN 1006) is 
$\theta(1006) = 30.60 \pm 0.06$ arcmin whereas that of Tycho is $\theta(Tycho) 
= 8.65 \pm 0.05$ arcmin. We can
 place ${{\theta(Tycho)}\ / {\theta(1006)}}\ =\ {{d(1006)}\ / {d(Tycho)}}\ 
{{t(Tycho)}\ / {t(1006)}}\ \eta$,
where $t(Tycho)$ and $t(1006)$ are the respective ages of the remnants and
$\eta = v(Tycho)/v(1006)$ is the ratio of the average expansion velocities 
of the two SN ejecta.  Such approach gives, if we use 
the distance obtained by Winkler, Gupta \& Long (2003) to SN 1006, 
a distance to SN 1572 of $d(Tycho) \simeq 3.3$ kpc, 
well in the middle of other estimates.

All of the above shows that it is at present 
hard to escape from the uncertain range
in the distance to SN 1572, i.e. from 2 to 4 kpc with the analyses
done so far. Such limitation does not prevent to infer  the 
right place of SN 1572 within the luminosity sample: the
conclusion on the brightness--rate of decline for SN 1572 is based on
the gathered empirical evidence and the well--established use of the
peak luminosity--decline correlation.

Incidentally, if a good distance would be available for SN 1572, one could
obtain the distance to SN 1996X, which is not well determined with
the present available methods. The D$_{n}$--$\sigma$ relation gives a
distance modulus $\mu$ = 31.32 $\pm$ 0.4 (Faber et al. 1989), whereas 
the SNeIa calibration using H$_{0}$ = 65 km s$^{-1}$ Mpc$^{-1}$ would
place it a significantly larger distance (Salvo et al. 2001).
 The very good agreement between the 
late time light curves of SN 1572 and SN 1996X allows
us to infer a $\Delta\mu$(SN1572--SN1996X) $=$ 18.9 $\pm$ 0.12 between both
SNeIa.

\section{Conclusions}

We have found that SN 1572 was a supernova very close to the template
with a {\it stretch factor}  s $\sim$ 0.9 (see Table 3 for a summary). 
The light curve grows in precision towards the late 
times, being highly uncertain around maximum brightness.
An overall agreement between early, late decline and color
with the expected evolution of normal SNeIa 
supports our conclusion.

Type Ia supernovae with {\it stretch factors} between 0.9 and 1.1 
make the vast majority of the observed population. They are not only
those most frequently found in nearby searches, but also the bulk
of discoveries in cosmological searches at high--z, as can be seen in
the sample of SNe at z $>$ 0.3 found by the 
Supernova Cosmology Project (Perlmutter et al. 1999). Among SNeIa
of s $\sim$ 0.9 in nearby galaxies 
for which very late--time data are available, we have  
found a close resemblance to SN 1996X in rate 
of decline. SN 1572 likely has a slightly slower rate. However, whereas 
SN 1996X was not heavily reddened, the reddening in SN 1572 is
E(B--V)=0.60 $\pm$ 0.05.  
While at present heavily reddened SNe Ia involve a larger difficulty
than unreddened ones 
for their proper use in cosmology (Riess et al. 2000; Knop et al. 2003), 
a better use of multicolor light curves should 
allow to overcome the problem  in their calibration for cosmology. 

Tycho Brahe's supernova belongs to the SNIa type that are used 
for the determination of the nature of dark energy.
Its light, traveling already light years away from us, carries the
information on our acceleration epoch to an even more dark--energy
dominated future. If used for cosmology by observers in galaxies
billions of light years distant, it would require, certainly, a
careful handling of its intrinsic brightness.

\bigskip

\noindent{\bf Acknowledgements} 
This paper amongst others to come is dedicated to my father. Past, present
and future merged for him just a few months ago, as sadly as
unexpectedly. His life was devoted to help the souls of many and to
carry research in brain and mind studies. His inner light lives forever
with us. I would like to express my gratitude to  
M. A. Granada, from the Department of History of Philosophy of the
U. Barcelona, for helping to locate the XVI century historical sources.

\vfill\eject

{}

\clearpage

\bigskip

\begin{table*}[htb]
\caption{Visual estimates of Tycho's Supernova}
\label{table:1}
\newcommand{\m}{\hphantom{$-$}}
\newcommand{\cc}[1]{\multicolumn{1}{c}{#1}}
\renewcommand{\tabcolsep}{0.05pc} 
\begin{tabular}{@{}llllll}
\hline
\hline
Date     &    &Phase  & Description      &   &   \\
 & (days)& Adopted &  & $m_{V}$ & Ref. \\
\hline
         &   &   &                         &   &        \\

1572 Nov  2 & Nov 2  & - & No detection  &$ >$ 5 & $^{1}$  \\
1572 Nov   & Nov 11 &  -10 & Somewhat brighter than  &
-3.$\pm$0.2 & $^{2}$  \\ 
         &  &  &  Jupiter  &      &     \\
         &  &  & and almost equal to the Morning Star &      &    \\
1572 Nov   & Nov 11 &  -10 & Equalled Venus when this  &
-3.$\pm$0.2 & \\ 
         &  &  & planet was at maximum brightness  &      &  $^{3}$  \\

1572 Nov 16,17  & Nov 16 $\pm$ 1 & -5 $\pm$ 1 & Almost as bright as
Venus&
 -4.0$\pm$0.3 & $^{4}$  \\
         &  &  & and Jupiter  &      &    \\
         &  &  &                            &         &  \\
1572 Dec  & Dec 15 $\pm$ 7 & 24$\pm$ 7  & About as bright as Jupiter
  &  -2.4  $\pm$ 0.3  & $^{6}$ \\
1573 Jan 7 & Jan 7  & 47 & Already fainter than Jupiter&
$>$-2. & $^{7}$ \\
1573 Jan & Jan 15$\pm$ 7 & 55 $\pm$ 7 & A little fainter than Jupiter &
-1.4 $\pm$ 0.4 & $^{7}$  \\

         &  &    & and brighter than the brighter 
  &   &   \\

         &  & & stars of first mag
  &   &   \\

1573 Feb--Mar &  Mar 2  $\pm$ 14& 101$\pm$14 & Equal to brighter stars
          of first mag &+0.3 $\pm$  0.2 
          & $^{8}$ \\
1573 Apr--May &  May 1$\pm$ 14 &161 $\pm$14  & Equal to stars of
          second mag. & +1.6 $\pm$ 0.5  
          & $^{9}$ \\
1573 July--Aug & Aug 1$\pm$ 14 & 253 $\pm$14 & Equal to $\alpha$, $\beta$, 
$\gamma$, $\delta$ Cas& +2.5 $\pm$ 0.2 & $^{10}$ \\
         &  &  &                            &        &  \\
1573 Oct--Nov & Nov 1$\pm$ 14 & 345$\pm$14 & Equal to stars of fourth
          mag. & +4.0 $\pm$ 0.2 & $^{11}$ \\
         &   &   &  (among Cassiopea stars)             &         & \\
1573 Nov & Nov 15 $\pm$ 7 & 359 $\pm$7 & Equal to $\kappa$ Cas &
          +4.2  $\pm$0.2 &$^{12}$  \\
         &   &   &                         &         & \\
1573 Dec--1574 Jan & Jan 1 $\pm$ 14&  406$\pm$ 14 &  Hardly exceeding stars of fifth mag. &
  +4.7  $\pm$ 0.2&  $^{13}$ \\
         &   &   &                         &         & \\
1574 Feb & Feb 15 $\pm$ 7 &451 $\pm$ 7 & Equal to stars of sixth
          mag. & +5.3 $\pm$ 0.4 & $^{13}$ \\
         &   &   &                         &         & \\
1574 Mar & Mar 15 $\pm$ 7 & 479$\pm$ 7 &  Invisible & $>$ 6 & $^{14}$ \\
         &   &   &                         &         & \\

\hline 
\hline
\end{tabular}\\[2pt]
\end{table*}

\newpage

\noindent
{{\bf NOTES TO TABLE 1} 

\bigskip

\noindent
$^{1}$ Jer\'onimo Mu\~noz writes that he was certain that on 
November 2, 1572 the ``comet'' was not in the sky, as he was
{\it teaching to know the stars to a numerous group of people that evening}
(Mu\~noz 1573).\footnote{  
Jer\'onimo Mu\~noz (?-1592) was professor of Hebrew and
Mathematics at the University of Valencia and later on of {\it Astrology}
at the University of Salamanca. The records assembled here are from his 
Book of the New Comet ({\it Libro del nuevo cometa})  
written as King Philip II consulted him on SN 1572.
He determines the absence of parallax
of the star and gives estimates of color and magnitudes, though
his account is mostly addressed to criticize  the Aristotelian 
cosmology. He wrote that the heavens and stars are not made of quintessence, 
but related to the elements, and advocated to reconsider 
the presocratic cosmological views of Democritus
and Anaxagoras. Tycho Brahe (1603b) compiled parts of Mu\~noz's observations
and discussed those views in the third volume of the {\it Progymnasmata}.
He left out, however, some useful information that we have added here. We
corrected as well the premaximum quote.}

\bigskip

\noindent
$^{2}$ Nov 11 is the first observation quoted by Mu\~noz in his book. 
The apparent magnitude appeared {\it somewhat greater than that of
Jupiter, which was 59 degrees distant from the comet, and
almost equal to that of the Morning Star}. Jupiter was at that
time V=--2.61 and Venus V=--4.35 (Baade 1945). We assign --3.0$\pm$0.3.

\bigskip

\noindent 
$^{3}$
On Nov 11 Tycho Brahe also saw it, and his preliminar report
in {\it De nova stella} said that {\it equaled Venus when 
this planet was at maximum brightness}. He was, however, more
conservative in the account he gives  in {\it Progymnasmata}. 
We give --3.5 mag to the estimate of Tycho $\pm$0.3. 
In the following we must keep in mind that Tycho Brahe observes 
from a latitude of $\sim$ N55$^{0}$, as he stays in the area of Copenhague
for two years since the discovery of SN 1572. From that location the
supernova can be observed well above the horizon through the whole year. 
De Vaucouleurs notes that when comparing  SN 1572 with Venus
one has to account from the fact that the earlier is not 
affected by extinction while the second was significantly affected. 
This places V=--3.8 or fainter
if one considers the apparent magnitude of Venus
affected by extinction.

\bigskip

\noindent 
$^{4}$
Tycho Brahe reports in the {\it Progymnasmata}
on the observations done by Caspar Peucer and Johannes 
Pr\"atorious on Nov 16. 
According to the first, {\it it surpassed in brightness
all stars and planets with the exception of Venus}, for the second,
it was {\it brighter than Jupiter but easily fainter than Venus}. 
Michael M\"astlin who saw it before Nov 17 states that 
{\it almost equalled Venus}. It has been given to these
estimates a value of V=--4.0 $\pm$0.3. 

\smallskip

\noindent
$^{5}$ Tycho Brahe does not give the exact date of the observations.
He quotes the month or an interval encompassing two months. It is 
reasonable to think that he refers to dates in the middle of the 
period. Therefore, we follow here Baade (1945) and assign the middle
of the period as date of observation. However, we add $\pm$ 7 days of 
uncertainty in the case a month is quoted, and $\pm$ 14 days if the
period encompasses two months. 

\smallskip

\noindent
$^{6}$ This is an upper limit given by Mu\~noz. 
He writes, {\it when it began to be visible, it looked larger
than Jupiter, and now, on January 7, 1573, it already looks smaller
than Jupiter; this could have happened because it had risen higher 
than where it was when it first appeared}. This upper limit to the
rate of decline as well as the data around it by Tycho Brahe constrain
the light curve stretch factor below {\it s} $=$ 1.1.

\smallskip
\noindent
$^{7}$ This is a difficult estimate. The interval quoted 
encompasses 2.46 mag (see Baade 1945).
 It is uncertain the interpolation done, from
Jupiter with V=--2.18 to  the brightest stars of 
the first class which are +0.28.  We assign to the Baade estimate
of V=--1.4 an uncertainty of $\pm$ 0.4.  

\smallskip
\noindent
$^{8}$ This a bit better estimate, as encompasses an smaller
range in magnitudes $\pm$ 0.2 if one selects the brighter range 
of first magnitude stars (0.2--0.3).  

\smallskip
\noindent
$^{9}$ It is a very ambiguous quote, one should give it 
 $\pm$ 0.5. The stars of second magnitude had a range encompassing
1 mag, $m_{V} =$4.0 and  $\sigma_{V} =$0.5. 

This is the point that lies far from the expected curve. 
It has been checked whether a problem of differential 
extinction could have affected SN 1572 with respect to the second
magnitude stars, but there were second magnitude stars at similar
altitude as SN 1572. The offset of 0.6 mag can not be attributed
to extinction problems, but to ambiguous description of the record.

\smallskip
\noindent
$^{10}$ Tycho Brahe's estimates in magnitude
are more precise now as he refers to particular stars not far
from the supernova with a low scatter in magnitude.
The mean of those stars is $m_{V} =$2.48 and  $\sigma_{V} =$0.23.
We assign $\pm$0.2 to the error in the mags of those stars.

\smallskip
\noindent
$^{11}$ It becomes a very precise estimate by comparing to the 
fourth mag stars in Cassiopeia, when knowing that it was brighter than
the nearby
eleventh star of Cassiopeia. An error of 0.2 mag is assigned.

\smallskip
\noindent
$^{12}$ This seems to be the most precise description as it says it is
equal to $\kappa$ Cas.

\smallskip
\noindent
$^{13}$ Errors of 0.2 and 0.4 mag respectively 
seem reasonable for these last estimates.

\smallskip
\noindent
$^{14}$ It can be noted that the quote that it was not
observable later than March 1574 (V $>$ 6.0) allows to exclude  
that the luminosity levelled off, due to a light echo, at 
$\sim$ 10 mag below the supernova maximum brightness. 

}

\clearpage

\bigskip

\begin{table*}[htb]
\caption{Colors of Tycho's Supernova}
\label{table:1}
\newcommand{\m}{\hphantom{$-$}}
\newcommand{\cc}[1]{\multicolumn{1}{c}{#1}}
\renewcommand{\tabcolsep}{0.2pc} 
\begin{tabular}{@{}llllll}
\hline
\hline
Date     &   & Phase  & Description      & B -- V  & Observer  \\
    &   & Adopted &    & Adopted & Reference \\
\hline
         &   &   &                         &   &        \\

1572 Nov  & Nov 21 $\pm$7  & -6 $\pm$ 7 & Like Venus (B -- V = 0.83)
& 0.82$\pm$0.25 &  $^{1}$  \\
         &  &  & and Jupiter (B -- V = 0.82)&      &    \\

1572 Dec & Dec 1 $\pm$ 5  & 10 $\pm$ 5 & Yellowish   &
         1.0 $\pm$ 0.25      & $^{2}$ \\

1572 Nov-Dec & Dec 2$_{-22}^{7}$ & 11$_{-22}^{7}$ & Between Saturn 
(B -- V = 1.04) & 1.2$\pm$ 0.25 & $^{3}$ \\ 
         &   &  & and Mars (B--V)=1.36,     &
&  \\
         &   &  & closer to Mars  &         &  \\

1572 Jan & Jan 1 $\pm$ 7  & 41 $\pm$ 7  & Like Aldebaran       
      (B--V=1.52)      &   1.52$\pm$ 0.25  & $^{4}$ \\

1573 Jan  & Jan 15 $\pm$7 &55 $\pm$ 7 & Similar to Mars (B -- V =
1.36)& 1.36 $\pm$0.2 & $^{5}$ \\

         &   &   &      &   &  \\
1573 Feb   &  Feb 28 $\pm$7  & 99$\pm$7 & Return to original &  0.82$\pm$0.25                   &   $^{6}$      \\
         &   &  &                            &         \\

1573 May  & May 15 $\pm$7 &175 $\pm$ 7 & Like the original (B -- V =
0.83)&0.82 $\pm$0.25 & $^{7}$
         \\
         &   &  &                            &         & \\
\hline
\end{tabular}\\[2pt]
\end{table*}

\noindent
{\bf NOTES TO TABLE 2} 

\smallskip

\noindent
$^{1}$Various reports (see Baade 1945) give a color similar to the colors of
Venus and Jupiter. 

\smallskip

\noindent
$^{2}$ Tycho Brahe (1603a) states that after maximum it became yellowish in
color. By comparison to the description as whitish (like Jupiter) 
and red like Aldebaran, we assign B--V $=$1.0 $\pm$0.25.  

\noindent
$^{3}$ Mu\~noz (1573) states that the color
was between those of Saturn and Mars, closer to Mars. 
One could identify then the color
as 1.2 $\pm$ 0.1.  Unfortunately, there is ambiguity in the date he is
refering to: 
he mentions that he first saw it
on Dec 2, and checked with observers that it was already there on Nov
11. He uses the expression ``the color then was'', so it
is not clear whether he refers to Nov 11 or Dec 2. 
It seems that the description, because of its detail,
is based on his own appreciation of color on Dec 2
(we place an errorbar in the dates
towards Nov 11, and a 7 days error after Dec 2). 

\smallskip

\noindent
$^{4}$Tycho Brahe notes that shortly after the 
beginning of the year it was red
like Aldebaran (Brahe 1603a). We assign 0.25 mag of
 uncertainty to those estimates. 

\smallskip

\noindent
$^{5}$ According to Pr\"atorius it was like Mars two months after its
appearance (around Jan 15). 

\smallskip

\noindent
$^{6}$M\"astlin refers that at the end of February it is going back 
to its whitish color.
Tycho states that it had a red color along February and March, 
becoming whitish after May. However, this seems to be a worse 
account of the color than the one given by M\"astlin. In April it is 
described as silverish by some observers in Spain, and as similar
to Saturn (van den Bergh 1970).

\smallskip

\noindent
$^{7}$ The last estimate
is given 0.82 $\pm$0.25 in account for Tycho Brahe on the
color in May. 

\smallskip

\clearpage

\begin{table*}[htb]
\caption{SN 1572}
\smallskip
\label{table:1}
\newcommand{\m}{\hphantom{$-$}}
\newcommand{\cc}[1]{\multicolumn{1}{c}{#1}}
\renewcommand{\tabcolsep}{0.5pc} 
\begin{tabular}{@{}llllll}
\hline
\hline
Parameter      &   {\it s}  &  ${\Delta} m(V)_{20}$  &  ${\Delta} m(V)_{60}$
 &  {\it ${\gamma}_{(V)}^{\it late}$} &
  {\it $B-V_{0}$} \\
    &   &   &   &    &  \\
\hline
         &   &    &  &   & \\

SN 1572  &  0.9 $\pm$ 0.05     & 1.1 $\pm$0.4   & 2.6 $\pm$0.6 
 & 1.6 $\pm$0.04  & 0.22
         $\pm$ 0.25 \\
         &  &   &    &  & \\

SN 1991bg--like  &  0.62  &  1.68 &  3.3 & 2.7 &   0.6  \\
         &  &   &    &   & \\

SN 1991T--like  &  1.2   &  0.72 & 2.46  & 1.68 &   -0.1  \\
         &  &   &    &    & \\

Normal$^{1}$  &  0.9 &  0.9  &  2.8  &  1.7 &   0.0  \\
         &  &   &     &   & \\

\hline
\end{tabular}\\[2pt]
\end{table*}

\noindent{Normal SNe Ia of s = 0.9.
Whereas the ${\Delta} m(V)_{20}$ is not well defined for SN 1572 due
to a lack of accurate observations in the week around the maximum, the
rate of decline in terms of stretch can be well determined by 
fitting premaximum and postmaximum up to 60 days.}

\newpage

\begin{figure}[hbtp]
\input epsf
\centerline{\epsfysize16cm\epsfbox{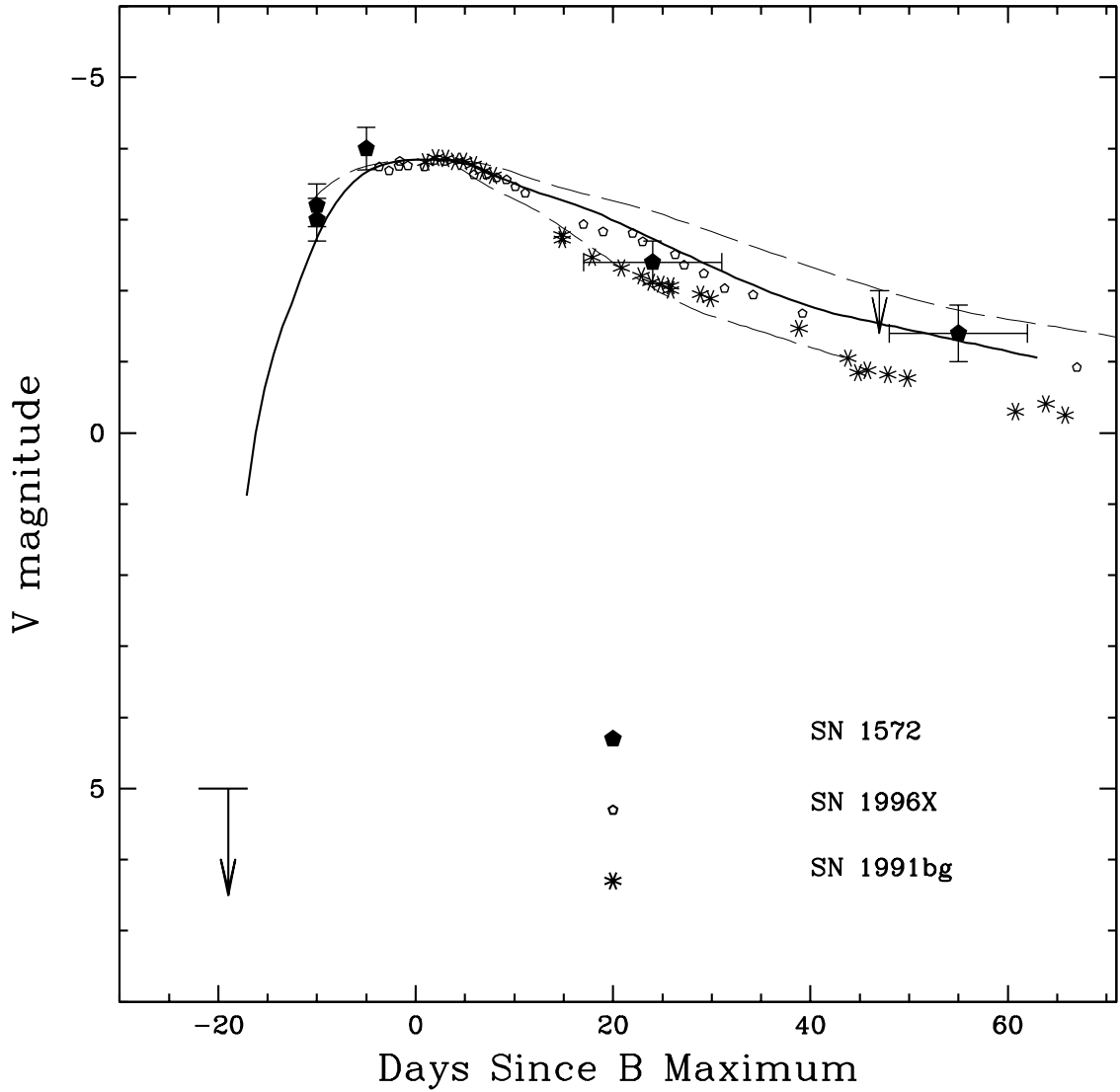}}
\nopagebreak[4]

\caption{The visual light curve of SN 1572 till 60 days. The solid curve
is the V light curve of a s $=$ 0.9 SNIa, which gives the best account
for the decline. Such {\it stretch factor} is typical of normal
SNe Ia. We show for comparison the V light curve of the
 normal SNIa SN 1996X, whose
stretch factor is {\it s} $=$ 0.889 and of the fast--declining 
SN 1991bg. SN 1572 was significantly slower than SN 1991bg.
The light curves plotted in dashed lines are the templates of
91bg--type events and 91T--like events, wich depart significantly from
the data.}
\label{fig2}
\end{figure}

\begin{figure}[hbtp]
\input epsf
\centerline{\epsfysize16cm\epsfbox{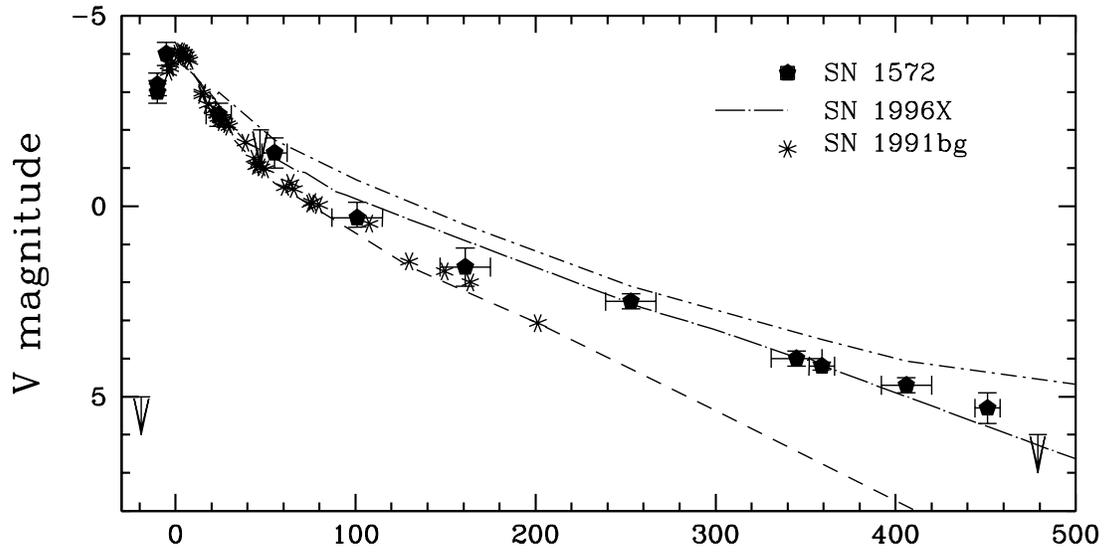}}
\nopagebreak[4]

\caption{The visual light curve of SN 1572 till 500 days. 
Its late rate of decline is the one of normal SNeIa. It is
very similar to the decline of the s $=$ 0.889 SN 1996X. The 
visual data of SN1991bg and the template light curves of this SNIa
and SN1991T are shown for comparison. 
}
\label{fig2}
\end{figure}

\begin{figure}[hbtp]
\input epsf
\centerline{\epsfysize16cm\epsfbox{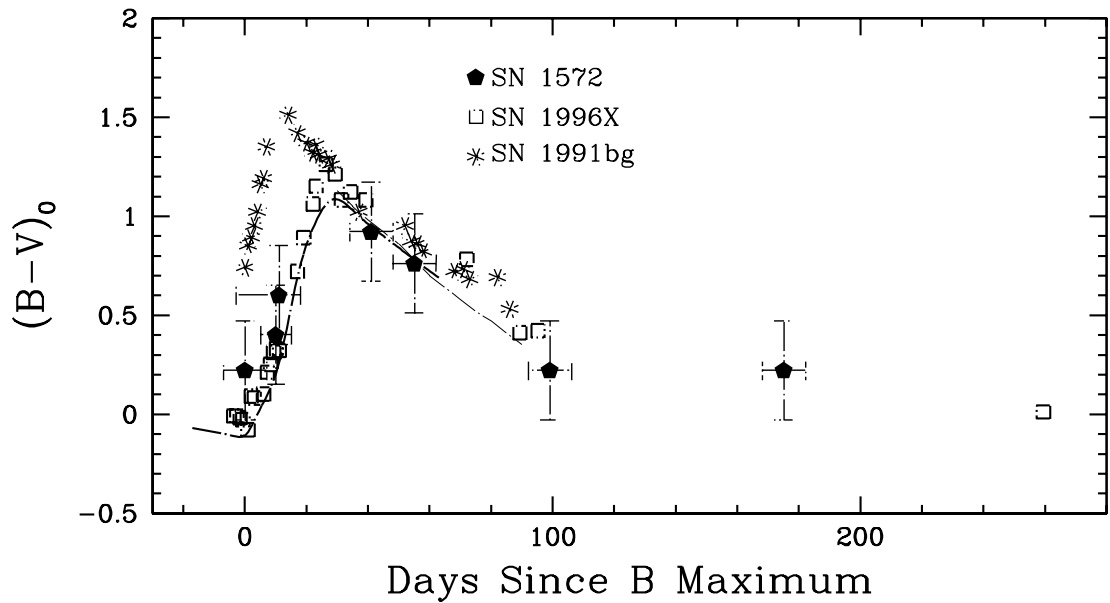}}
\nopagebreak[4]

\caption{
Color evolution of SN 1572 corrected from extinction as
compared with normal SNe Ia and SN 1991bg. SN 1572 is consistent
with the color evolution of a s $=$ 0.9 SNIa (template for this stretch
is plotted as dot--dashed curve). SN 1996X has been 
corrected for its very small
reddening E(B-V)=0.01 $\pm$0.02
 as well as SN 1991bg (E(B--V)=0.03 $\pm$0.05) (Ph99).}
\label{fig2}
\end{figure}

\vfill\eject

\end{document}